\documentclass[a4paper,10pt,doublesided, doublecolumn,final]{IEEEtran}  
\IEEEoverridecommandlockouts
\usepackage{float}
\usepackage{algorithm}
\usepackage[noend]{algorithmic}
\usepackage[geometry]{ifsym}
\usepackage {ifsym}
\usepackage{amssymb}
\usepackage{yhmath}
\usepackage{amscd}
\usepackage{dsfont}
\usepackage{amsmath}
\usepackage{epsfig}
\usepackage{graphics}
\usepackage{psfrag}
\usepackage{rotating}
\usepackage{amsmath}
\usepackage{amsfonts}
\usepackage{url}
\usepackage{color}
\usepackage{float}
\usepackage{balance} 
\usepackage{soul}

\newtheorem{lemma}{Lemma}

\newtheorem{theorem}{Theorem}
\newtheorem{definition}{Definition}

\newcommand{\ind}[1]{\mathds{1}_{\left\lbrace #1 \right\rbrace}}

\newcommand{\bs}{\boldsymbol}

\newcommand{\ds}{\displaystyle}

\newcommand{\sfT}{\textsf{T}}

\newcommand{\Cldicnof}{\mathcal{C}}
\newcommand{\Nldicnof}{\mathcal{N}_{\eta}}

\newcommand{\Uldicnfb}{ U }

\newcommand{\Bldicnof}{\mathcal{B}_{\eta}}

\newcommand{\GameNF}{\mathcal{G} = \left(\mathcal{K}, \left\lbrace\mathcal{A}_k \right\rbrace_{k \in \mathcal{K}},\left\lbrace
u_{k}\right\rbrace_{ k \in \mathcal{K}}\right)}

 \renewcommand{\baselinestretch}{0.97}

\begin{document}
\title{Nash Region of the Linear Deterministic Interference Channel with Noisy Output Feedback}
\author{Victor Quintero, Samir M. Perlaza, Jean-Marie Gorce, and H. Vincent Poor
\thanks{Victor Quintero, Samir M. Perlaza and Jean-Marie Gorce  are with the Laboratoire CITI (a joint laboratory between the Universit\'e de Lyon, INRIA, and INSA de Lyon). 6 Avenue des Arts,  F-69621, Villeurbanne,  France. ($\lbrace$victor.quintero-florez, samir.perlaza, jean-marie.gorce$\rbrace$@inria.fr).}
\thanks{H. Vincent Poor is with the Department of Electrical Engineering at Princeton University, Princeton, NJ 08544 USA. (poor@princeton.edu).}
\thanks{Victor Quintero is also with Universidad del Cauca, Popay\'an, Colombia.}
\thanks{Samir M. Perlaza is also with the Department of Electrical Engineering at Princeton University, Princeton, NJ 08544 USA.}
\thanks{This research was supported in part by the European Commission under Marie Sk\l{}odowska-Curie Individual Fellowship No. 659316; in part by the INSA Lyon - SPIE ICS chair on the Internet of Things; in part by the Administrative Department of Science, Technology, and Innovation of Colombia (Colciencias), fellowship No. 617-2013; and in part by the U. S. National Science Foundation under Grants CCF-1420575  and ECCS-1343210.}
}
\maketitle
\begin{abstract}  
In this paper, the $\eta$-Nash equilibrium ($\eta$-NE) region of the two-user linear deterministic interference channel (IC) with noisy channel-output feedback is characterized for all $\eta > 0$. The $\eta$-NE region, a subset of the capacity region, contains the set of all achievable information rate pairs that are stable in the sense of an $\eta$-NE. More specifically, given an $\eta$-NE coding scheme, there does not exist an alternative coding scheme for either transmitter-receiver pair that increases the individual rate by more than $\eta$ bits per channel use. Existing results such as the $\eta$-NE region of the linear deterministic IC without feedback and with perfect output feedback are obtained as particular cases of the result presented in this paper.
\end{abstract}

\section{System Model} \label{SecChModelLDICNOF}

\begin{figure}[t!]
 \centerline{\epsfig{figure=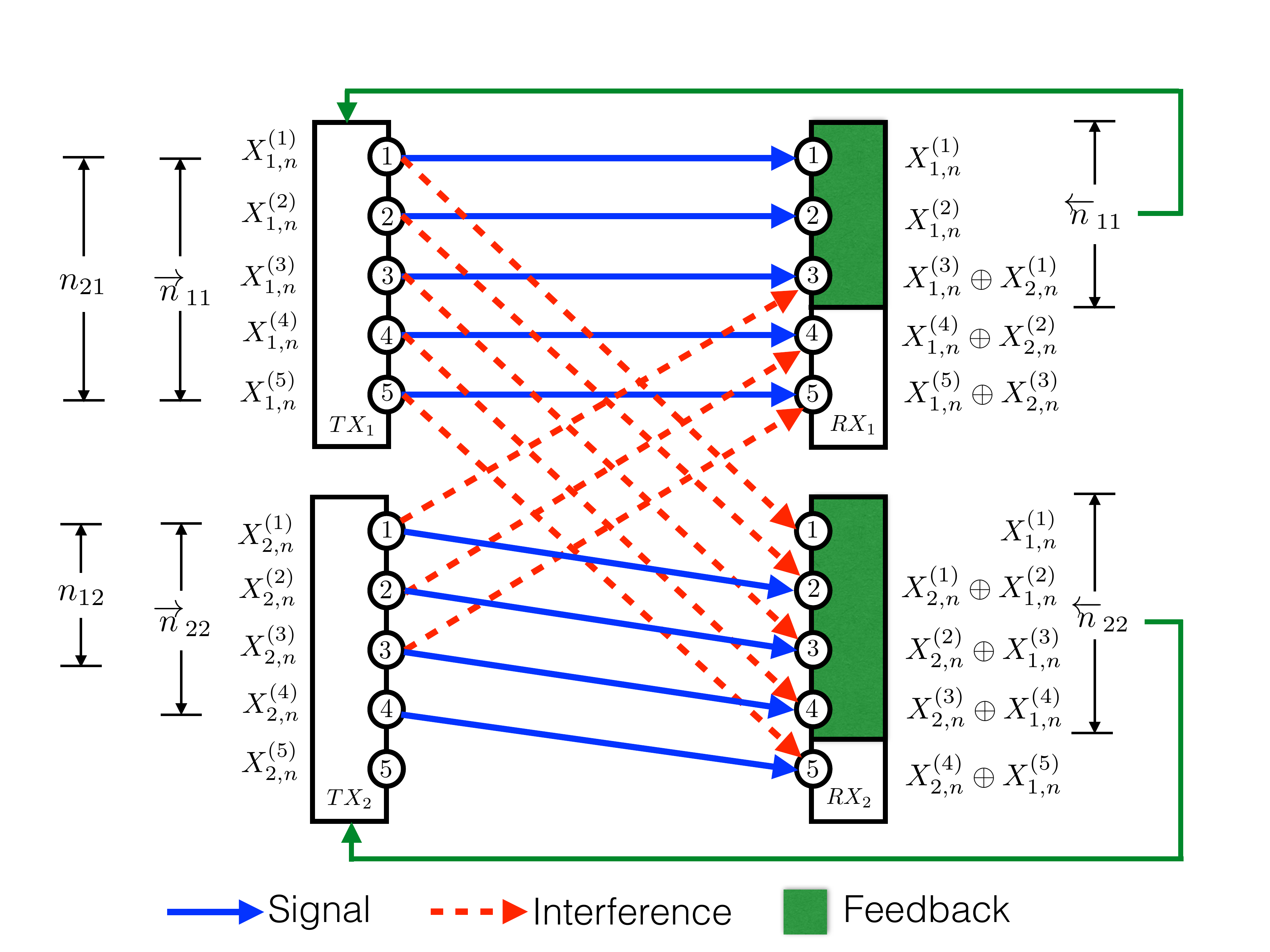,width=0.5\textwidth}}
 \vspace{-0.1in}
\caption{Two-user linear deterministic interference channel with noisy channel-output feedback at channel use $n$.} 
  \label{FigLD-IC-NOF}
\end{figure}

Consider the two-user decentralized linear deterministic interference channel with noisy channel-output feedback (D-LD-IC-NOF) depicted in Figure \ref{FigLD-IC-NOF}. 
For all $i \in \lbrace 1, 2 \rbrace$, with $j\in \lbrace 1, 2 \rbrace\setminus \lbrace i \rbrace$, the number of bit-pipes between transmitter $i$ and its intended receiver  is denoted by $\overrightarrow{n}_{ii}$; the number of bit-pipes between transmitter $i$ and its non-intended receiver is denoted by $n_{ji}$; and the number of bit-pipes between receiver $i$ and its corresponding transmitter is denoted by  $\overleftarrow{n}_{ii}$. These six non-negative integer parameters describe the D-LD-IC-NOF in Figure \ref{FigLD-IC-NOF}.

At transmitter $i$, the channel-input $\bs{X}_{i,n}$ at channel use $n$, with $n \in \lbrace 1, 2,  \ldots, N_i \rbrace$, is a $q$-dimensional binary vector ${\bs{X}_{i,n} = \left(X_{i,n}^{(1)}, X_{i,n}^{(2)}, \ldots, X_{i,n}^{(q)}\right)^{\sfT}} \in \mathcal{X}_i$, with $\mathcal{X}_i = \lbrace 0, 1 \rbrace^{q}$,   
\begin{equation}\label{Eqq}
q=\ds\max \left(\overrightarrow{n}_{11}, \overrightarrow{n}_{22}, n_{12}, n_{21}\right),
\end{equation}
and $N_i \in \mathds{N}$ is the block-length of transmitter-receiver pair $i$. 
At receiver $i$, the channel-output $\overrightarrow{\bs{Y}}_{i,n}$ at channel use $n$, with $n \in \lbrace 1, 2,  \ldots, \max\left(N_1,N_2\right) \rbrace$, is also a $q$-dimensional binary vector ${\overrightarrow{\bs{Y}}_{i,n} = \left(\overrightarrow{Y}_{i,n}^{(1)}, \overrightarrow{Y}_{i,n}^{(2)}, \ldots, \overrightarrow{Y}_{i,n}^{(q)}\right)^{\sfT}}$. 
Let $\bs{S}$ be a $q\times q$ binary lower shift matrix. The input-output relation during channel use $n$ is given by 
\begin{IEEEeqnarray}{lcl}
\label{EqLDICsignals}
\overrightarrow{\bs{Y}}_{i,n}&=& \bs{S}^{q - \overrightarrow{n}_{ii}} \bs{X}_{i,n} + \bs{S}^{q - n_{ij}} \bs{X}_{j,n},
\end{IEEEeqnarray}
where ${\bs{X}_{i,n} = \left(0, 0, \ldots, 0\right)^{\sfT}}$ for all $n > N_i$.
The feedback signal $\bs{\overleftarrow{Y}}_{i,n}$ available at transmitter $i$ at the end of channel use $n$ is  
\begin{IEEEeqnarray}{lcl}\label{EqLDICsignalsc}
\bs{\overleftarrow{Y}}_{i,n} &=& \bs{S}^{\left(\max(\overrightarrow{n}_{ii},n_{ij})-\overleftarrow{n}_{ii}\right)^+} \, \bs{\overrightarrow{Y}}_{i,n-d}, \quad
\end{IEEEeqnarray}
where $d$ is a finite delay, additions and multiplications are defined over the binary field, and $(\cdot)^+$ is the positive part operator.

Without any loss of generality, the feedback delay is assumed to be equal to one channel use. 
Let $\mathcal{W}_i$ be the set of message indices of transmitter $i$.
Transmitter $i$ sends the message index $W_i \in \mathcal{W}_i$ by transmitting the codeword ${\bs{X}_{i}=\left(\bs{X}_{i,1}, \bs{X}_{i,2}, \ldots, \bs{X}_{i,N_i}\right) \in \mathcal{X}_{i}^{N_i}}$, which is a binary $q \times N_i$ matrix. 
The encoder of transmitter $i$ can be modeled as a set of deterministic mappings $f_{i,1}^{(N)}, f_{i,2}^{(N)}, \ldots, f_{i,N_i}^{(N)}$, with $f_{i,1}^{(N)}: \mathcal{W}_i \times \mathds{N} \rightarrow \lbrace 0, 1 \rbrace^{q}$ and for all $n \in \lbrace 2, 3, \ldots, N_i\rbrace$, $f_{i,n}^{(N)}: \mathcal{W}_i \times \mathds{N}  \times \lbrace 0, 1 \rbrace^{q \times (n-1)} \rightarrow \lbrace 0, 1 \rbrace^{q}$, such that 
\begin{subequations}
\label{EqEnconderf}
\begin{IEEEeqnarray}{lcl}
\bs{X}_{i,1} &=& f_{i,1}^{(N)}\big(W_i, \Omega_i \big) \mbox{ and }\\
\bs{X}_{i,n} &=& f_{i,n}^{(N)}\big(W_i, \Omega_i, \bs{\overleftarrow{Y}}_{i,1}, \bs{\overleftarrow{Y}}_{i,2}, \ldots, \bs{\overleftarrow{Y}}_{i,n-1} \big), 
\end{IEEEeqnarray}
\end{subequations}
where $\Omega_i$ is a randomly generated index known by both transmitter $i$ and receiver $i$, while unknown by transmitter $j$ and receiver $j$.
The decoder of receiver $i$ is defined by a deterministic function $\psi_i^{(N)}: \lbrace 0,1 \rbrace^{q \times N} \times \mathds{N} \rightarrow \mathcal{W}_i$.
At the end of the communication, receiver $i$ uses the $q \times N$ binary matrix  $\left(\overrightarrow{\bs{Y}}_{i,1}, \overrightarrow{\bs{Y}}_{i,2}, \ldots, \overrightarrow{\bs{Y}}_{i,N}\right)$ and $\Omega_i$ to obtain an estimate $\widehat{W}_i \in \mathcal{W}_i$ of the message index $W_i$, i.e., $\widehat{W}_i=\psi_i^{(N)}\left(\overrightarrow{\bs{Y}}_{i,1}, \overrightarrow{\bs{Y}}_{i,2}, \ldots, \overrightarrow{\bs{Y}}_{i,N}, \Omega_i \right)$.  
Let $W_i$ be written as $c_{i,1} \, c_{i,2} \, \ldots \, c_{i,M_i}$ in binary form, with $M_i=\lceil \log_{2} | \mathcal{W}_i | \rceil$. 
Let also $\widehat{W}_i$ be written as $\widehat{c}_{i,1} \, \widehat{c}_{i,2} \, \ldots \, \widehat{c}_{i,M_i}$ in binary form. 

A transmit-receive configuration for transmitter-receiver pair $i$, denoted by $s_i$, can be described in terms of the block-length $N_i$, the number of bits per block $M_i$, the channel-input alphabet $\mathcal{X}_i$,   the codebook, the encoding functions $f_{i,1}^{(N)}, f_{i,2}^{(N)}, \ldots, f_{i,N_i}^{(N)}$, the decoding function $\psi_i^{(N)}$, etc.

The average bit error probability at decoder $i$ given the configurations $s_1$ and $s_2$, denoted by $p_{i}(s_1,s_2)$, is given by
\begin{IEEEeqnarray}{rcl}
\label{EqbitErrorProb}   
p_{i}(s_1,s_2)   &=& \frac{1}{M_i} \ds\sum_{\ell = 1}^{M_i} \ind{\widehat{c}_{i,\ell} \neq {c}_{i,\ell} }. 
\end{IEEEeqnarray}
Within this context, a rate pair $(R_1,R_2) \in \mathds{R}_+^{2}$ is said to be achievable if it complies with the following definition. 
\begin{definition}[Achievable Rate Pairs]\label{DefAchievableRatePairs}\emph{
A rate pair $(R_1,R_2) \in \mathds{R}_+^{2}$ is achievable if there exists at least one pair of configurations $(s_1,s_2)$ such that the decoding bit error probabilities $p_{1}(s_1,s_2)$ and $p_{2}(s_1,s_2)$ can be made arbitrarily small by letting the block-lengths $N_1$ and $N_2$ grow to infinity.}
\end{definition}

The aim of transmitter $i$ is to autonomously choose its transmit-receive configuration $s_i$, in order to maximize its achievable rate $R_i$.
Note that the rate achieved by transmitter-receiver $i$ depends on both configurations $s_1$ and $s_2$ due to mutual interference. This reveals the competitive interaction between both links in the decentralized interference channel. 
The following section models this interaction using tools from game theory.

\section{The Two-User Interference Channel as a Game}\label{SecGameFormulation}

The competitive interaction between the two transmitter-receiver pairs in the decentralized interference channel can be modeled by the following game in normal-form:
\begin{equation}\label{EqGame}
\GameNF.
\end{equation}
The set $\mathcal{K} = \lbrace 1, 2 \rbrace$ is the set of players, that is, the set of transmitter-receiver pairs. The sets $\mathcal{A}_1$ and $\mathcal{A}_2$ are the sets of actions of players $1$ and $2$, respectively. An action of a player $i \in \mathcal{K}$, which is denoted by $s_i \in \mathcal{A}_i$, is basically its transmit-receive configuration as described in Section \ref{SecChModelLDICNOF}. 
The utility function of player $i$ is $u_i: \mathcal{A}_1 \times \mathcal{A}_2 \rightarrow \mathds{R}_+$ and it is defined  as the information rate of transmitter $i$,
\begin{equation}
\label{EqUtility}
u_i(s_1,s_2) = \left\lbrace
\begin{array}{lcl}
R_i=\frac{M_i}{N_i}, & \mbox{if} &  p_{i}(s_1,s_2)  < \epsilon \\
0, & &  \mbox{otherwise,}
\end{array}
\right.
\end{equation}
where $\epsilon > 0$ is an arbitrarily small number. 

This game formulation was first proposed in \cite{Yates-ISIT-2008} and \cite{Berry-ISIT-2008}. A class of transmit-receive configurations $\bs{s}^* = (s_1^*, s_2^*) \in \mathcal{A}_1 \times \mathcal{A}_2$ that are particularly important in the analysis of this game is referred to as the set of $\eta$-Nash equilibria ($\eta$-NE), with $\eta>0$. This type of configuration satisfies the following definition. 
\begin{definition}[$\eta$-Nash equilibrium] \label{DefEtaNE} \emph{
In the game ${\GameNF}$, an action profile  $(s_1^*, s_2^*)$ is an $\eta$-Nash equilibrium if for all $i \in \mathcal{K}$ and for all $s_i \in \mathcal{A}_i$, there exits an $\eta > 0$ such that
\begin{equation}\label{EqNashEquilibrium}
u_i (s_i , s_j^*) \leqslant u_i (s_i^*, s_j^*) + \eta.
\end{equation}
}
\end{definition}

Let $(s_1^*, s_2^*)$ be an $\eta$-Nash equilibrium action profile of the game in \eqref{EqGame}. Then, none of the transmitters can increase its own information transmission rate more than $\eta$ bits per channel use by changing its own transmit-receive configuration and keeping the average bit error probability arbitrarily close to zero. 
Note that for $\eta$ sufficiently large, from Definition  \ref{DefEtaNE}, any pair of configurations can be an $\eta$-NE. Alternatively, for $\eta=0$, the classical definition of Nash equilibrium is obtained \cite{Nash-PNAS-1950}. In this case, if a pair of configurations is a Nash equilibrium ($\eta=0$), then each individual configuration is optimal with respect to each other. Hence, the interest is to describe the set of all possible $\eta$-NE rate pairs $(R_1,R_2)$ of the game in \eqref{EqGame} with the smallest $\eta$ for which there exists at least one equilibrium configuration pair.
The set of rate pairs that can be achieved at an $\eta$-NE is known as the $\eta$-Nash equilibrium region.
\begin{definition}[$\eta$-NE Region] \label{DefNERegion} \emph{
Let $\eta>0$ be fixed. An achievable rate pair $(R_1,R_2)$ is said to be in the $\eta$-NE region of the game $\GameNF$ if there exists a pair $(s_1^*, s_2^*) \in \mathcal{A}_1 \times \mathcal{A}_2$  that is  an  $\eta$-NE and the following holds:
\begin{eqnarray}
u_1 (s_1^* , s_2^*)  =  R_1 & \mbox{ and } & u_2 (s_1^* , s_2^*)  =  R_2. 
\end{eqnarray}
}
\end{definition}

The following section characterizes the $\eta$-NE region (Def. \ref{DefNERegion})  of the two-user D-LD-IC-NOF in \eqref{EqGame}, denoted by $\Nldicnof\left(\overrightarrow{n}_{11}, \overrightarrow{n}_{22}, n_{12}, n_{21}, \overleftarrow{n}_{11}, \overleftarrow{n}_{22}\right)$,  for fixed parameters  $\big(\overrightarrow{n}_{11}$, $\overrightarrow{n}_{22}$, $n_{12}$, $n_{21}$, $\overleftarrow{n}_{11}$,$\overleftarrow{n}_{22}\big)\in \mathds{N}^6$ and for all $\eta>0$.  

\section{Main Results}\label{SectMainResultsLDICNOF}
\begin{figure*}[ht!] 
\centerline{\epsfig{figure=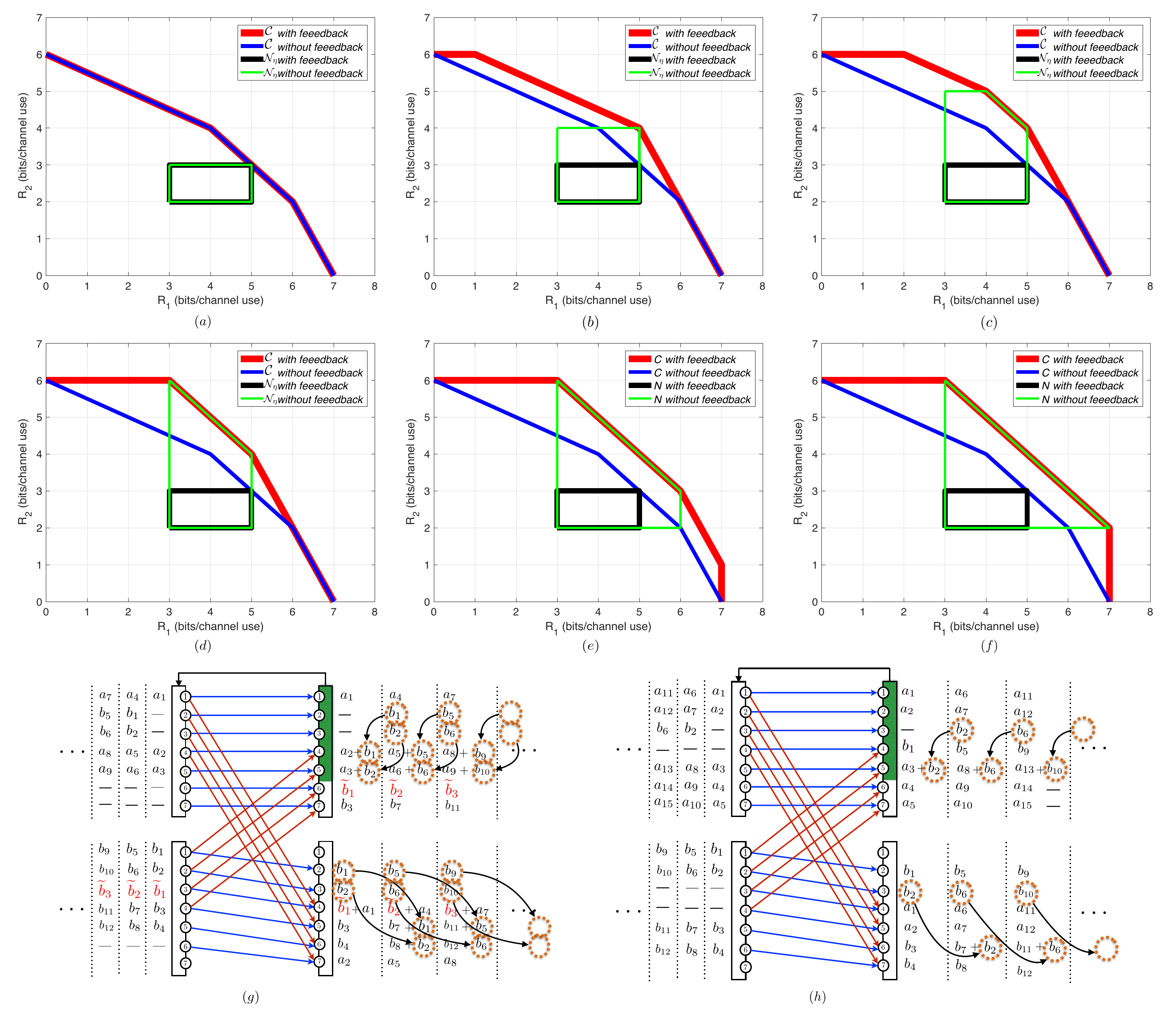,width=0.85\textwidth}}
\vspace{-0.2in}
 \caption{Capacity region $\Cldicnof(7,6,4,4,0,0)$  (thin blue line) and $\eta$-NE region $\Nldicnof(7,6,4,4,0,0)$ (thick black line) with $\eta$ chosen arbitrarily small. Fig. \ref{FigRegionLDM-N}a  shows the capacity region $\Cldicnof(7,6,4,4,\protect \overleftarrow{n}_{11}, \protect \overleftarrow{n}_{22})$ (thick red line) and the $\eta$-NE region $\Nldicnof(7,6,4,4,\protect \overleftarrow{n}_{11}, \protect \overleftarrow{n}_{22})$ (thin green line), with $\protect \overleftarrow{n}_{11} \in \lbrace 0,1,2,3,4 \rbrace$ and $\protect \overleftarrow{n}_{22} \in \lbrace 0,1,2,3,4 \rbrace$. Fig. \ref{FigRegionLDM-N}b  shows the capacity region $\Cldicnof(7,6,4,4,5, \protect \overleftarrow{n}_{22})$ (thick red line) and the $\eta$-NE region $\Nldicnof(7,6,4,4,5, \protect \overleftarrow{n}_{22})$ (thin green line), with $\protect \overleftarrow{n}_{22} \in \lbrace 0,1,2,3,4 \rbrace$. Fig. \ref{FigRegionLDM-N}c  shows the capacity region $\Cldicnof(7,6,4,4,6, \protect \overleftarrow{n}_{22})$ (thick red line) and the $\eta$-NE region $\Nldicnof(7,6,4,4,6, \protect \overleftarrow{n}_{22})$ (thin green line), with $\protect \overleftarrow{n}_{22} \in \lbrace 0,1,2,3,4 \rbrace$. Fig. \ref{FigRegionLDM-N}d  shows the capacity region $\Cldicnof(7,6,4,4,7, \protect \overleftarrow{n}_{22})$ (thick red line) and the $\eta$-NE region $\Nldicnof(7,6,4,4,7, \protect \overleftarrow{n}_{22})$ (thin green line), with $\protect \overleftarrow{n}_{22} \in \lbrace 0,1,2,3,4 \rbrace$. Fig. \ref{FigRegionLDM-N}e  shows the capacity region $\Cldicnof(7,6,4,4,7,5)$ (thick red line) and the $\eta$-NE region $\Nldicnof(7,6,4,4,7, 5)$ (thin green line). Fig. \ref{FigRegionLDM-N}f  shows the capacity region $\Cldicnof(7,6,4,4,7,6)$ (thick red line) and the $\eta$-NE region $\Nldicnof(7,6,4,4,7, 6)$ (thin green line). Fig. \ref{FigRegionLDM-N}g and Fig. \ref{FigRegionLDM-N}h  illustrate the achievability scheme for the equilibrium rate pair $(3,4)$ and $(5,4)$ in $\Nldicnof{(7,6,4,4,5,0)}$.}
 \label{FigRegionLDM-N}
\vspace{-0.5cm}
\end{figure*}

The $\eta$-NE region is characterized in terms of two regions: the capacity region, denoted by $\Cldicnof{\left(\overrightarrow{n}_{11}, \overrightarrow{n}_{22}, n_{12}, n_{21}, \overleftarrow{n}_{11}, \overleftarrow{n}_{22}\right)}$ and a convex region, denoted by $\Bldicnof{\left(\overrightarrow{n}_{11}, \overrightarrow{n}_{22}, n_{12}, n_{21}, \overleftarrow{n}_{11}, \overleftarrow{n}_{22}\right)}$. 
In the following, the tuple  $(\overrightarrow{n}_{11}$, $\overrightarrow{n}_{22}$, $n_{12}$, $n_{21}$, $\overleftarrow{n}_{11}$, $\overleftarrow{n}_{22})$ is used only when needed. 

The capacity region $\Cldicnof$ of the two-user LD-IC-NOF is described in Theorem $1$ in \cite{QPEG-TIT-2016}, which is a generalization of previous works in \cite{Bresler-ETT-2008} and \cite{Suh-TIT-2011}.
For all $\eta > 0$, the convex region $\Bldicnof$ is defined as follows:
\begin{IEEEeqnarray}{lcl}
\label{EqBldicnfb}
\Bldicnof &=& \Big \lbrace (R_1, R_2): \! L_i \! \leqslant \! R_i \! \leqslant \! \Uldicnfb_i, \! \mbox{ for all } i \in \lbrace 1, 2 \rbrace \Big \rbrace, 
\end{IEEEeqnarray}
where, 
\begin{subequations}
\label{EqLlUidicnfb}
\begin{IEEEeqnarray}{lcl}
\label{EqLldicnfb}
&L_i&  =  \left(\left(\overrightarrow{n}_{ii} - n_{ij} \right)^+-\eta\right)^+ \mbox{ and } \\
\label{EqUldicnfb}
&U_i&  =   \max\left( \overrightarrow{n}_{ii}, n_{ij} \right) - \Bigg(\min\left(\left(\overrightarrow{n}_{jj} \! - \! n_{ji} \right)^+, n_{ij}\right) \\
\nonumber
& \! - \! & \! \bigg( \! \min \! \left( \! \left( \! \overrightarrow{n}_{jj}\! - \! n_{ij} \! \right)^+ \!, \! n_{ji} \! \right) \! - \! \left( \! \max \! \left( \! \overrightarrow{n}_{jj}, \! n_{ji} \! \right) \! - \! \overleftarrow{n}_{jj} \! \right)^+ \!  \bigg)^+ \! \Bigg)^+\!+\! \eta,
\end{IEEEeqnarray}
\end{subequations}
with $i\in \lbrace 1, 2 \rbrace$ and $j\in \lbrace 1, 2 \rbrace\setminus \lbrace i \rbrace$.
Theorem \ref{TheoremMainResultLDICnFB} uses the region $\Bldicnof$ in \eqref{EqBldicnfb} and the capacity region $\Cldicnof$ to describe the $\eta$-NE region $\Nldicnof$. 
\begin{theorem}\label{TheoremMainResultLDICnFB}\emph{
Let $\eta > 0$ be fixed. The $\eta$-NE region $\Nldicnof$ of the two-user D-LD-IC-NOF with parameters $\overrightarrow{n}_{11}$, $\overrightarrow{n}_{22}$, $n_{12}$, $n_{21}$, $\overleftarrow{n}_{11}$ and $\overleftarrow{n}_{22}$, is $\Nldicnof =  \Cldicnof \cap \Bldicnof$.
}
\end{theorem}
Figure \ref{FigRegionLDM-N} shows the capacity region $\Cldicnof$ and the $\eta$-NE region $\Nldicnof$ of a channel with parameters $\overrightarrow{n}_{11} = 7$, $\overrightarrow{n}_{22} = 6$, $n_{12} = 4$, $n_{21} = 4$ and different values for  $\overleftarrow{n}_{11}$ and $\overleftarrow{n}_{22}$, with $\eta$ chosen arbitrarily small.  Note that when $\overleftarrow{n}_{11} \in \lbrace 0,1,2,3,4\rbrace$ and $\overleftarrow{n}_{22} \in \lbrace 0,1,2,3,4\rbrace$ (Figure \ref{FigRegionLDM-N}a), it follows that  $\Nldicnof{(7,6,4,4,\overleftarrow{n}_{11},\overleftarrow{n}_{22})} = \Nldicnof{(7,6,4,4,0,0)}$. Thus, in this case the use of feedback in any of the transmitter-receiver pairs does not enlarge the $\eta$-Nash region. Alternatively, when $\overleftarrow{n}_{11}>4$ and $\overleftarrow{n}_{22} \in \lbrace 0,1,2,3,4\rbrace$ (Figures \ref{FigRegionLDM-N}b, \ref{FigRegionLDM-N}c and \ref{FigRegionLDM-N}d), the resulting $\eta$-Nash region is strictly larger than in the previous case. A similar effect is observed in Figures \ref{FigRegionLDM-N}e and \ref{FigRegionLDM-N}f. This observation implies the existence of a threshold on each feedback parameter $\overleftarrow{n}_{11}$ and $\overleftarrow{n}_{22}$ beyond which the $\eta$-Nash region is enlarged. The exact values of $\overleftarrow{n}_{11}$ and $\overleftarrow{n}_{22}$, given a fixed tuple $(\overrightarrow{n}_{11}$, $\overrightarrow{n}_{22}$, $n_{12}$, $n_{21})$, beyond which the $\eta$-Nash region can be enlarged is presented in \cite{Quintero-INRIA-RR-2017}. 
Figure \ref{FigRegionLDM-N}g and Figure \ref{FigRegionLDM-N}h show the coding schemes to achieve the rate pairs $(3,4)$ and $(5,4)$, respectively,  when $\overleftarrow{n}_{11}=5$ and $\overleftarrow{n}_{22}=0$. In Figure \ref{FigRegionLDM-N}g, note that common randomness is used by transmitter-receiver pair $2$ to prevent transmitter-receiver pair $1$ from  increasing its individual rate.  More specifically, the bits $\tilde{b}_1$, $\tilde{b}_2$, $\tilde{b}_3$, \ldots are known by both transmitter $2$ and receiver $2$. The use of common randomness is also observed in \cite{Berry-TIT-2011, Perlaza-TIT-2015} and \cite{Perlaza-ISIT-2014a}. Common randomness reflects a competitive behavior between both transmitter-receiver pairs. In Figure \ref{FigRegionLDM-N}g, common randomness is not used by transmitter-receiver pair $2$ and thus, transmitter-receiver pair $1$ achieves a higher rate at an $\eta$-NE with respect to the previous example. This suggests a more altruistic behavior. 

The $\eta$-NE region $\Nldicnof$ without feedback, i.e., when $\overleftarrow{n}_{11} = 0$ and $\overleftarrow{n}_{22} = 0$ (Theorem $1$ in \cite{Berry-TIT-2011}), is $\Nldicnof{(\overrightarrow{n}_{11}, \overrightarrow{n}_{22}, n_{12}, n_{21}, 0, 0)}$.
The $\eta$-NE region with perfect feedback i.e., $\overleftarrow{n}_{11} = \max(\overrightarrow{n}_{11}, n_{12})$ and $\overleftarrow{n}_{22} = \max(\overrightarrow{n}_{22}, n_{21})$ (Theorem $1$ in \cite{Perlaza-TIT-2015}),  is $\Nldicnof(\overrightarrow{n}_{11}$, $\overrightarrow{n}_{22}$, $n_{12}$, $n_{21}$, $\max(\overrightarrow{n}_{11}, n_{12})$, $\max(\overrightarrow{n}_{22}, n_{21}))$.
From the comments above, it is interesting to highlight the following inclusions:
\begin{IEEEeqnarray} {ll}
\label{Eqsetinclussions}
\Nldicnof &\Big(\overrightarrow{n}_{11}, \overrightarrow{n}_{22}, n_{12}, n_{21}, 0, 0\Big) \subseteq \\
\nonumber
\Nldicnof & \Big(\overrightarrow{n}_{11}, \overrightarrow{n}_{22}, n_{12}, n_{21}, \overleftarrow{n}_{11}, \overleftarrow{n}_{22}\Big) \subseteq \\
\nonumber
\Nldicnof & \Big(\overrightarrow{n}_{11}, \overrightarrow{n}_{22}, n_{12}, n_{21}, \max\left( \overrightarrow{n}_{11},n_{12}\right), \max\left( \overrightarrow{n}_{22},n_{21}\right)\Big),
\end{IEEEeqnarray} 
for all $\eta > 0$. The inclusions above might appear trivial, however, enlarging the set of actions often leads to paradoxes (Braess Paradox \cite{Braess-U-1969}) in which the new game possesses equilibria at which players obtain smaller individual benefits and/or smaller total benefit. Nonetheless, letting both transmitter-receiver pairs to use feedback does not induce this type of paradoxes with respect to the case without feedback.  

\section{Proofs} \label{SecProofs-LD-IC-NOF}
To prove Theorem \ref{TheoremMainResultLDICnFB}, the first step is to show that a rate pair $(R_1, R_2)$, with $R_i <  L_i$ or $R_i >  U_i$ for at least one $i \in \lbrace 1, 2 \rbrace$, is not achievable at an $\eta$-equilibrium for all $\eta>0$. That is, 
\begin{equation}
\label{EqTheorem1Part1}
\Nldicnof \subseteq \Cldicnof  \cap \Bldicnof.
\end{equation}
The second step is to show that, for all $\eta>0$, any point in $\Cldicnof \cap \Bldicnof$ can be achievable at an $\eta$-equilibrium. That is, 
\begin{equation}
\label{EqTheorem1Part2}
\Nldicnof \supseteq \Cldicnof \cap \Bldicnof,
\end{equation}
which proves the equality $\Nldicnof = \Cldicnof \cap \Bldicnof$.

\paragraph{Proof of \eqref{EqTheorem1Part1}} 
The proof of \eqref{EqTheorem1Part1} is completed by the following lemmas.
\begin{lemma} \label{LemmaProofTheorem1part11} \emph{ A rate pair $(R_1,R_2) \in \Cldicnof$, with either $R_1 < L_1$ or $R_2 < L_2$ is not achievable at an $\eta$-equilibrium for all $\eta > 0$.
}
\end{lemma}
\begin{IEEEproof}
The proof of Lemma \ref{LemmaProofTheorem1part11} is presented in \cite{Quintero-INRIA-RR-2017}. 
\end{IEEEproof}
The intuition behind this proof is that the rate ${R_i = \left(\overrightarrow{n}_{ii}-n_{ij}\right)^+}$ is always achievable independently of the coding scheme of transmitter-receiver pair $j$. To achieve ${R_i = \left(\overrightarrow{n}_{ii}-n_{ij}\right)^+}$ transmitter $i$ uses the most significant bit-pipes, which are interference free, to transmit new bits at each channel use $n$.
\begin{lemma} \label{LemmaProofTheorem1part12} \emph{ A rate pair $(R_1,R_2) \in \Cldicnof$, with either $R_1 > U_1$ or $R_2 > U_2$ is not achievable at an $\eta$-equilibrium for all $\eta > 0$.
}
\end{lemma}
\begin{IEEEproof}
The proof of Lemma \ref{LemmaProofTheorem1part12} is presented in \cite{Quintero-INRIA-RR-2017}.
\end{IEEEproof}
This proof is based on the fact that at an $\eta$-NE, transmitter $j$ might re-transmit some of the bits previously transmitted by transmitter $i$. The interference produced by those re-transmitted bits  at receiver $i$ can be eliminated if they were received interference free during previous channel uses. This allows transmitter $i$ to use the bit-pipes interfered with by those re-transmitted bits to send new information bits at each channel use. The key point of this proof is to show that the maximum number of bits that can be re-transmitted at an $\eta$-NE is upper bounded.

\paragraph{Proof of \eqref{EqTheorem1Part2}} 
Consider a modification of the coding scheme with noisy feedback presented in \cite{QPEG-TIT-2016}, which combines rate splitting \cite{Han-TIT-1981}, block Markov superposition coding \cite{Cover-TIT-1981} and backward decoding \cite{Willems-PhD-1982}.  The novelty with respect to \cite{QPEG-TIT-2016} consists of allowing users to introduce common randomness as suggested in \cite{Berry-TIT-2011} and \cite{Perlaza-TIT-2015}. 

Consider without any loss of generality that $N=N_1=N_2$. Let $W_i^{(t)}  \in \lbrace1, 2,  \ldots, 2^{NR_i}\rbrace$ and  $\Omega_i^{(t)} \in \lbrace1, 2,  \ldots, 2^{NR_{i,R}}\rbrace$ denote the message index and the random message index sent by transmitter $i$ during the $t$-th block, with $t \in \lbrace 1, 2, \ldots, T\rbrace$, respectively. Following a rate-splitting argument, assume that $\left(W_i^{(t)}, \Omega_i^{(t)} \right)$ is represented by the indices $\left( W_{i,C1}^{(t)}, \Omega_{i,R1}^{(t)} , W_{i,C2}^{(t)}, \Omega_{i,R2}^{(t)}, W_{i,P}^{(t)}\right) \in \lbrace 1, 2,  \ldots, 2^{N R_{i,C1}} \rbrace \times \lbrace 1, 2,  \ldots, 2^{NR_{i,R1}} \rbrace \times \lbrace 1, 2,  \ldots, 2^{N R_{i,C2}} \rbrace \times \lbrace 1, 2,  \ldots, 2^{N R_{i,R2}} \rbrace \times \lbrace 1, 2, \ldots, 2^{N R_{i,P}} \rbrace$, where $R_{i}= R_{i,C1} + R_{i,C2}+R_{i,P} $ and $R_{i,R}=R_{i,R1} + R_{i,R2}$.
The rate $R_{i,R}$ is the number of transmitted bits that are known by both transmitter $i$ and receiver $i$ per channel use, and thus it does not have an impact on the information rate $R_i$.

The codeword generation follows a four-level superposition coding scheme.
The indices  $W_{i,C1}^{(t-1)}$ and $\Omega_{i,R1}^{(t-1)}$ are assumed to be decoded at transmitter $j$ via the feedback link of transmitter-receiver pair $j$ at the end of the transmission of block $t-1$. Therefore, at the beginning of block $t$, each transmitter possesses the knowledge of the indices $W_{1,C1}^{(t-1)}$, $\Omega_{1,R1}^{(t-1)}$, $W_{2,C1}^{(t-1)}$ and $\Omega_{2,R1}^{(t-1)}$. In the case of the first block $t = 1$, the indices  $W_{1,C1}^{(0)}$, $\Omega_{1,R1}^{(0)}$, $W_{2,C1}^{(0)}$ and $\Omega_{1,R2}^{(0)}$ are assumed to be known by all transmitters and receivers.
Using these indices both transmitters are able to identify the same codeword in the first code-layer. This first code-layer, which is common for both transmitter-receiver pairs, is a sub-codebook of $2^{N\left(R_{1,C1} + R_{2,C1}+R_{1,R1} + R_{2,R1}\right)}$ codewords. Denote by $\bs{u}\left(W_{1,C1}^{(t-1)}, \Omega_{1,R1}^{(t-1)}, W_{2,C1}^{(t-1)}, \Omega_{2,R1}^{(t-1)}\right)$ the corresponding codeword in the first code-layer. 
The second codeword is chosen by transmitter $i$ using  $\left(W_{i,C1}^{(t)},\Omega_{i,R1}^{(t)}\right)$ from the second code-layer, which is a sub-codebook of $2^{N\left(R_{i,C1}+R_{i,R1}\right)}$ codewords corresponding to the codeword $\bs{u}\left(W_{1,C1}^{(t-1)}, \Omega_{1,R1}^{(t-1)}, W_{2,C1}^{(t-1)}, \Omega_{2,R1}^{(t-1)}\right)$. Denote by $\bs{u}_i\left(W_{1,C1}^{(t-1)}, \Omega_{1,R1}^{(t-1)}, W_{2,C1}^{(t-1)}, \Omega_{2,R1}^{(t-1)}, W_{i,C1}^{(t)}, \Omega_{i,R1}^{(t)}\right)$ the corresponding codeword in the second code-layer.  
The third codeword is chosen by transmitter $i$ using  $\left(W_{i,C2}^{(t)},\Omega_{i,R2}^{(t)}\right)$  from the third code-layer, which is a sub-codebook of $2^{N\left(R_{i,C2}+R_{i,R2}\right)}$ codewords corresponding to the codeword $\bs{u}_i \left( W_{1,C1}^{(t-1)}, \Omega_{1,R1}^{(t-1)}, W_{2,C1}^{(t-1)}, \Omega_{2,R1}^{(t-1)}, W_{i,C1}^{(t)}, \Omega_{i,R1}^{(t)} \right)$. Denote by $\bs{v}_i\Big(W_{1,C1}^{(t-1)}, \Omega_{1,R1}^{(t-1)}$, $W_{2,C1}^{(t-1)}, \Omega_{2,R1}^{(t-1)}$, $W_{i,C1}^{(t)}, \Omega_{i,R1}^{(t)}$, $W_{i,C2}^{(t)}, \Omega_{i,R2}^{(t)}\Big)$ the corresponding codeword in the third code-layer.   
The fourth codeword is chosen by transmitter $i$ using  $W_{i,P}^{(t)}$ from the fourth code-layer, which is a sub-codebook of $2^{N\,R_{i,P}}$ codewords corresponding to the codeword $\bs{v}_i\Big(W_{1,C1}^{(t-1)}, \Omega_{1,R1}^{(t-1)}$, $W_{2,C1}^{(t-1)}, \Omega_{2,R1}^{(t-1)}$, $W_{i,C1}^{(t)}, \Omega_{i,R1}^{(t)}$, $W_{i,C2}^{(t)}, \Omega_{i,R2}^{(t)}\Big)$. Denote by $\bs{x}_{i,P}\Big(W_{1,C1}^{(t-1)}, \Omega_{1,R1}^{(t-1)}$, $W_{2,C1}^{(t-1)}, \Omega_{2,R1}^{(t-1)}$, $W_{i,C1}^{(t)}, \Omega_{i,R1}^{(t)}$, $W_{i,C2}^{(t)}, \Omega_{i,R2}^{(t)}, W_{i,P}^{(t)}\Big)$ the corresponding codeword in the fourth code-layer. 
Finally, the codeword $\bs{x}_i\Big(W_{1,C1}^{(t-1)}, \Omega_{1,R1}^{(t-1)}$, $W_{2,C1}^{(t-1)}, \Omega_{2,R1}^{(t-1)}$, $W_{i,C1}^{(t)}, \Omega_{i,R1}^{(t)}$, $W_{i,C2}^{(t)}, \Omega_{i,R2}^{(t)}, W_{i,P}^{(t)}\Big)$ to be sent during block $t \in \lbrace 1, 2,  \ldots, T \rbrace$ is a simple concatenation of the previous codewords, i.e., ${\bs{x}_i = \left( \bs{u}_i^\sfT, \bs{v}_i^\sfT, \bs{x}_{i,P}^\sfT\right)^\sfT} \in \lbrace0,1\rbrace^{q \times N}$, where the message indices have been dropped for ease of notation. 

The decoder follows a backward decoding scheme.
In the following, this coding scheme is referred to as a randomized Han-Kobayashi coding scheme with noisy feedback (R-HK-NOF) and it is described in \cite{Quintero-INRIA-RR-2017}.
The rest of the proof consists of showing that the R-HK-NOF coding scheme is capable of achieving an $\eta$-NE with $(R_1,R_2) \in \Cldicnof \cap \Bldicnof$ for all $\eta>0$, subject to a proper choice of the rates $R_{i,R1}$ and $R_{i,R2}$, for all $i \in \lbrace 1,2\rbrace$. 

\begin{lemma}\label{LemmaLDICFBb}
\emph{ The achievable region of the randomized Han-Kobayashi coding scheme for the D-LD-IC-NOF is the set of non-negative rates $\Big(R_{1,C1}$, $R_{1,R1}$, $R_{1,C2}$, $R_{1,R2}$, $R_{1,P}$, $R_{2,C1}$, $R_{2,R1}$, $R_{2,C2}$, $R_{2,R2}$, $R_{2,P}\Big)$ that satisfy the following conditions for all $i \in \lbrace 1, 2 \rbrace$ and $j\in \lbrace 1, 2 \rbrace\setminus \lbrace i \rbrace$:
\begin{subequations}
\label{EqHK3}
\begin{IEEEeqnarray}{rcl} 
R_{j,C1}+R_{j,R1}  & \leqslant &  \theta_{1,i}, \\
R_{i} + R_{j,C}+ R_{j,R}  & \leqslant &  \theta_{2,i}, \\ 
R_{j,C2}+R_{j,R2}  & \leqslant & \theta_{3,i}, \\
R_{i,P}    &\leqslant &  \theta_{4,i}, \\
R_{i,P}+R_{j,C2}+R_{j,R2}  & \leqslant &  \theta_{5,i}, \\
R_{i,C2}+R_{i,P}  & \leqslant & \theta_{6,i},  \mbox{ and } \\
R_{i,C2}+R_{i,P}+R_{j,C2}+R_{j,R2} & \leqslant & \theta_{7,i},
\end{IEEEeqnarray}
\end{subequations}
where,
\begin{subequations}
\label{EqThetaLDM}
\begin{IEEEeqnarray}{rcl}
\theta_{1,i}&=& \left(n_{ij}-\left(\max\left(\overrightarrow{n}_{ii},n_{ij}\right)-\overleftarrow{n}_{ii}\right)^+\right)^+, \\
\theta_{2,i}&=&  \max\left(\overrightarrow{n}_{ii},n_{ij}\right), \\
\theta_{3,i}&=& \min \left(n_{ij},\left(\max\left(\overrightarrow{n}_{ii},n_{ij}\right)-\overleftarrow{n}_{ii}\right)^+\right), \\
\theta_{4,i}&=& \left(\overrightarrow{n}_{ii}-n_{ji}\right)^+,  \\
\nonumber
\theta_{5,i}&=&  \max \Big(\left(\overrightarrow{n}_{ii}-n_{ji}\right)^+,\\
& &\min\left(n_{ij},\left(\max\left(\overrightarrow{n}_{ii},n_{ij}\right)-\overleftarrow{n}_{ii}\right)^+\right)\Big), \\
\nonumber
\theta_{6,i}&=&  \min\left(n_{ji},\left(\max\left(\overrightarrow{n}_{jj},n_{ji}\right)-\overleftarrow{n}_{jj}\right)^+\right) \\
\nonumber
& & -\min\big(\left(n_{ji}-\overrightarrow{n}_{ii}\right)^+,\left(\max\left(\overrightarrow{n}_{jj},n_{ji}\right)-\overleftarrow{n}_{jj}\right)^+\big)\\
& &+\left(\overrightarrow{n}_{ii}-n_{ji}\right)^+,   \mbox{ and } \\
\nonumber
\theta_{7,i} &=&  \max\big(\min\big(n_{ij},\left(\max\left(\overrightarrow{n}_{ii},n_{ij}\right)-\overleftarrow{n}_{ii}\right)^+\big), \\
\nonumber
& & \min\big(n_{ji},\left(\max\left(\overrightarrow{n}_{jj},n_{ji}\right)-\overleftarrow{n}_{jj}\right)^+\big)\\
\nonumber
& & -\min\big(\left(n_{ji}-\overrightarrow{n}_{ii}\right)^+,\left(\max\left(\overrightarrow{n}_{jj},n_{ji}\right)-\overleftarrow{n}_{jj}\right)^+\big)\\
& & +\left(\overrightarrow{n}_{ii}-n_{ji} \right)^+\big).
\end{IEEEeqnarray}
\end{subequations}
 }
\end{lemma}
\balance
\begin{IEEEproof}
The proof of Lemma \ref{LemmaLDICFBb} is presented in \cite{Quintero-INRIA-RR-2017}.
\end{IEEEproof}

The set of inequalities in \eqref{EqHK3} can be written in terms of the transmission rates $R_1 = R_{1,C1}+R_{1,C2}+R_{1,P}$ and $R_2 = R_{2,C1}+R_{2,C2}+R_{2,P}$ to observe that the R-HK-NOF achieves all the rates $(R_1,R_2) \in \Cldicnof$, when $R_{1,R} = R_{2,R} = 0$. 

The following lemma shows than when both transmitter-receiver links use the R-HK-NOF scheme and one of them unilaterally changes its coding scheme, it obtains a rate improvement that can be upper bounded. 
\begin{lemma}\label{LemmaLDICFBc}
\emph{Let $\eta > 0$ be fixed and let the rate tuple $\bs{R} = (R_{1,C} - \frac{\eta}{6}, R_{1,R} - \frac{\eta}{6},R_{1,P}- \frac{\eta}{6},R_{2,C} - \frac{\eta}{6}, R_{2,R} - \frac{\eta}{6}, R_{2,P}- \frac{\eta}{6})$ be achievable with the R-HK-NOF such that $R_1 = R_{1,P} + R_{1,C} - \frac{1}{3}\eta$ and $R_2 = R_{2,P} + R_{2,C}- \frac{1}{3}\eta$. Then, any unilateral deviation of transmitter-receiver pair $i$ by using any other coding scheme leads to a transmission rate $R_i'$ that satisfies $R_i' \leqslant  \max\left(\overrightarrow{n}_{ii}, n_{ij} \right) - (R_{j,C} + R_{j,R}) +  \frac{2}{3}\eta$.
}
\end{lemma}
 \begin{IEEEproof} 
The proof of Lemma \ref{LemmaLDICFBc} is presented in \cite{Quintero-INRIA-RR-2017}.
\end{IEEEproof}
Lemma \ref{LemmaLDICFBc} reveals the relevance of the random symbols $\Omega_{1}$ and $\Omega_{2}$ used by the R-HK-NOF. 
Even though the random symbols used by transmitter $j$ do not increase the effective transmission rate of transmitter-receiver pair $j$, they strongly limit the rate improvement transmitter-receiver pair $i$ can obtain by deviating from the R-HK-NOF coding scheme.
This observation can be used to show that the R-HK-NOF can be  an $\eta$-NE, when both $R_{1,R}$ and $R_{2,R}$ are properly chosen.  
The following lemma formalizes this intuition.

\begin{lemma}\label{LemmaLDICFBd}
\emph{Let $\eta > 0$ be fixed and let the rate tuple $\bs{R} = (R_{1,C} - \frac{\eta}{6}, R_{1,R} - \frac{\eta}{6},R_{1,P} - \frac{\eta}{6},R_{2,C} - \frac{\eta}{6}, R_{2,R} - \frac{\eta}{6},R_{2,P}- \frac{\eta}{6})$ be achieved by using the R-HK-NOF, with
\begin{IEEEeqnarray}{rcl}
\label{EqConditionNEc}
R_{i,C}  + R_{i,P} + R_{j,C} + R_{j,R} & = & \max(\overrightarrow{n}_{ii},n_{ij}) + \frac{2}{3}\eta,
\end{IEEEeqnarray}
for all $i \in \lbrace 1, 2 \rbrace$. Then, the rate pair $(R_1,R_2)$, with $R_i = R_{i,C} + R_{i,P} - \frac{1}{3}\eta$ is achievable at an $\eta$-Nash equilibrium.
}
\end{lemma}

\begin{IEEEproof}
The proof of Lemma \ref{LemmaLDICFBd} is presented in \cite{Quintero-INRIA-RR-2017}.
\end{IEEEproof}
The following lemma shows that all the rate pairs $(R_{1}, R_{2}) \in \Cldicnof  \cap \Bldicnof$ are achievable by the R-HK-NOF coding scheme at an $\eta$-NE, for all $\eta>0$.
\begin{lemma}\label{LemmaLDICFBe}\emph{
Let $\eta > 0$ be fixed. Then, for all rate pairs $(R_{1}, R_{2}) \in  \Cldicnof  \cap \Bldicnof$, there always exists at least one $\eta$-NE transmit-receive configuration pair $(s_1^*,s_2^*) \in \mathcal{A}_1\times\mathcal{A}_2$, such that $u_1(s_1^*,s_2^*) = R_{1}$ and $u_2(s_1^*,s_2^*) = R_{2}$.
}
\end{lemma}

\begin{IEEEproof}
The proof of Lemma \ref{LemmaLDICFBe} is presented in \cite{Quintero-INRIA-RR-2017}.
\end{IEEEproof}
This proof consists of showing that the set of inequalities in \eqref{EqHK3} and \eqref{EqConditionNEc} leads to a set of rate pairs identical to  $\Cldicnof  \cap \Bldicnof$. 
This concludes the proof of Theorem~\ref{TheoremMainResultLDICnFB}.

\section{Conclusions}
In this paper, the $\eta$-NE region of the D-LD-IC-NOF has been characterized for all $\eta>0$. This region contains the  $\eta$-NE region without feedback studied in \cite{Berry-TIT-2011} and is contained within the $\eta$-NE region with perfect channel-output feedback studied in \cite{Perlaza-TIT-2015}.

\bibliographystyle{IEEEtran}
\bibliography{IT-GT}

\end{document}